\documentclass[aps,prmaterials,notitlepage,superscriptaddress,reprint]{revtex4-2}
\usepackage[final]{graphicx}
\usepackage{natbib}
\usepackage{hyperref}
\usepackage{amssymb}
\usepackage{amsmath}
\usepackage{mathrsfs}
\usepackage{xcolor}
\usepackage{bm}
\usepackage{comment}
\begin{document}

\newcommand{\LS}[1]{\textcolor{black}{#1}}

\title{Microscopic origin of polytype-dependent melting in SiC revealed by machine-learning molecular dynamics}
\author{Ljiljana Stojanovi\'c} 
\affiliation{Hartree Centre, STFC Daresbury Laboratory, Warrington WA4 4AD, United Kingdom}
\author{Samuel J. Magorrian}
\affiliation{Hartree Centre,  STFC Daresbury Laboratory, Warrington WA4 4AD, United Kingdom}
\author{Lara Kabalan} 
\affiliation{Hartree Centre,  STFC Daresbury Laboratory, Warrington WA4 4AD, United Kingdom}
\author{Richard N. White}
\affiliation{Lucideon Limited, Stone Business Park, Brooms Road, Stone, Staffordshire, ST15 0SH, UK}
\author{Fabian L. Thiemann} 
\affiliation{IBM Research Europe,
	Daresbury, WA4 4AD, UK}
\author{Viktor Z\'olyomi}
\affiliation{Hartree Centre, STFC Daresbury Laboratory, Warrington WA4 4AD, United Kingdom}

\begin{abstract}{Predicting how crystal structure influences high-temperature stability remains a key challenge in materials modelling and design. Silicon carbide (SiC), one of the most thermally and chemically stable materials known, provides an ideal system for studying this problem because its many polytypes preserve similar local tetrahedral bonding while differing in long-range stacking geometry. Here, we combine phase-coexistence machine-learning molecular dynamics with finite-temperature phonon analysis, enabled by a fine-tuned MACE interatomic potential that accurately describes crystalline, high-temperature, and disordered configurations across multiple SiC polytypes. We identify a clear relative stability ordering, $3\mathrm{C} > 2\mathrm{H} > 9\mathrm{R}$, reflected consistently in structural disordering, interlayer sliding, and finite-temperature phonon spectra. Across all polytypes, melting initiates through the formation of short C--C contacts and carbon-rich local regions, followed by a progressive loss of tetrahedral Si--C connectivity. The reduced stability of the long-period 9R polytype is traced to low-frequency transverse-acoustic shear modes associated with relative bilayer sliding, which are already present in the 0~K phonon spectra and soften further at high temperature. These modes generate larger lateral bilayer displacements, linking enhanced interlayer sliding to local chemical disordering and ultimately melting. More broadly, our results show that high-temperature stability in polytypic covalent materials is governed not only by local bond strength, but also by stacking-dependent transverse dynamics.
}
    
\end{abstract}

\maketitle

\section{Introduction}

\LS{Understanding how crystal structure affects high-temperature stability and melting is an important problem in condensed-matter physics and materials modelling. In the classical Born picture, melting is associated with a loss of shear rigidity, which can be reflected in the softening of transverse acoustic shear modes~\cite{Born1939}. In real materials, however, melting is a finite-temperature process involving anharmonic dynamics, local structural disorder, defect formation, and long length and time scales. This makes direct atomistic simulations challenging, especially for covalent materials with high melting temperatures. Machine-learning interatomic potentials (MLIPs) now make it possible to access these length and time scales with accuracy approaching that of first-principles methods, providing a route to connect lattice dynamics with microscopic disordering processes near melting.}

\LS{Polytypic materials are useful model systems for examining this connection. Different polytypes can have nearly identical local bonding environments while differing in medium- and long-range stacking order, symmetry, and phonon dispersions. Their relative high-temperature stability may therefore depend not only on local bond strength, but also on stacking-dependent lattice dynamics.}
\LS{Silicon carbide (SiC) is particularly well suited for this purpose. It is among the most thermally and chemically stable materials, with a high melting temperature, strong mechanical stability, radiation tolerance, and a wide band gap. These properties make SiC important for technologies operating under extreme conditions, including next-generation nuclear and fusion systems and high-temperature electronics~\cite{KATOH2019151849,SiC_fusion_2018,Sizyuk_2024}. SiC is also known for its polytypism, with more than 250 reported polytypes~\cite{cheung2006silicon}. These polytypes differ in the stacking sequence of close-packed Si--C bilayers while preserving the same first-neighbor tetrahedral coordination. Despite their similar local bonding environments, these polytypes exhibit measurable differences in electronic, mechanical, and thermodynamic properties~\cite{Park1994,Xu2018}. However, how subtle variations in stacking sequence influence their high-temperature stability remains less well understood.}

Experimentally, the melting temperature of SiC has been reported in the range of approximately 2800--3500~K, depending on measurement method, pressure, and interpretation~\cite{Bhaumik1996,Togaya1998,Ekimov2004,cryst8050217,Sokolov2012}. The nature of SiC melting also remains debated. Several studies report incongruent melting~\cite{Togaya1998,Ekimov2004,cryst8050217,Bhaumik1996}, whereas high-pressure experiments have led to different interpretations, including reports of congruent melting~\cite{Sokolov2012}. These uncertainties motivate atomistic simulations that can resolve microscopic disordering and melting dynamics. Previous computational studies have examined crystalline and amorphous SiC, high-temperature dynamics, phase transformations, and radiation response~\cite{Vashishta2007,Yan2013,Kubo2021,MacIsaac2024,Xie2023,Xie2026,Wu2025,LIU2024}. Recent machine-learning molecular dynamics simulations have provided new insight into SiC phase behaviour, including reconstruction of parts of the high-temperature phase diagram and the identification of carbon clustering as a microscopic precursor to destabilization in 3C-SiC~\cite{Xie2026,Wu2025}. However, these studies have primarily focused on the cubic 3C polytype, and the role of polytypism in SiC melting remains less explored.

Here, we perform a systematic investigation of melting in representative 3C, 2H, and 9R SiC polytypes (Figure \ref{fig:mace_DFPT}a) using phase-coexistence molecular dynamics enabled by a fine-tuned MACE equivariant machine-learning interatomic potential for SiC. This framework provides access to the system sizes and time scales required to resolve defect nucleation and strongly anharmonic dynamics near melting. By combining analyses of structural disorder, interlayer sliding, and finite-temperature phonon spectra obtained from velocity-current spectral analysis of MD trajectories, we examine how stacking-dependent lattice dynamics contributes to differences in high-temperature stability.

Across all polytypes, disordering is initiated by the formation of short C--C contacts and the progressive loss of tetrahedral Si--C connectivity. We identify a clear relative high-temperature stability ordering, $3\mathrm{C} > 2\mathrm{H} > 9\mathrm{R}$. This ordering is consistent with finite-temperature phonon spectra, which show that transverse-acoustic modes associated with shear-like sliding along the stacking direction have the lowest frequencies in 9R, intermediate frequencies in 2H, and the highest frequencies in 3C. The same trend is already present in the 0~K phonon spectra, indicating that stacking-dependent low-frequency modes provide a microscopic link between polytype geometry and relative high-temperature stability in covalent materials.

\section{Results}

\subsection{MACE Model Performance}

\begin{figure*}[h!]
    \centering
    \includegraphics[width=0.9\textwidth]{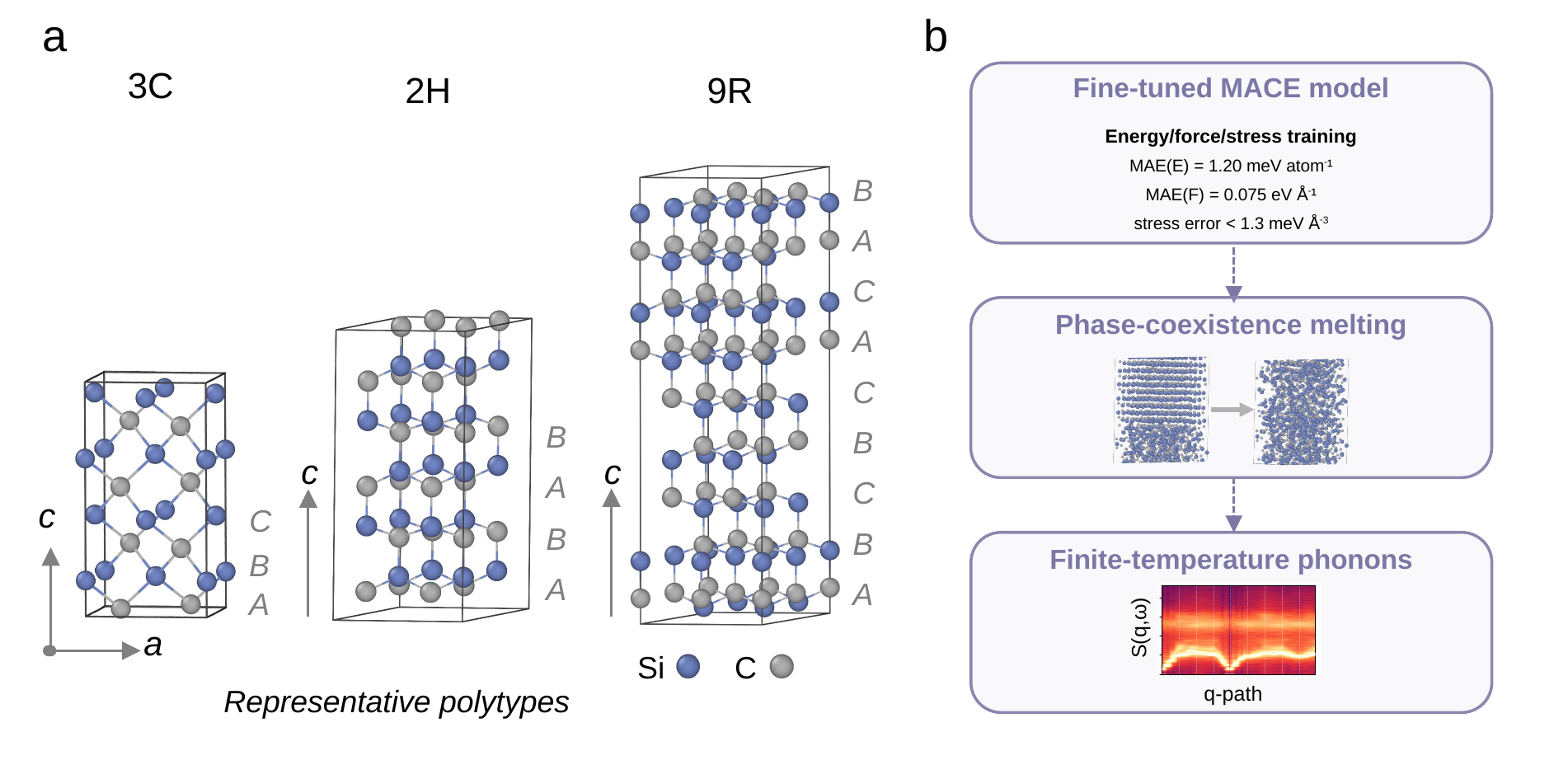}
    \caption{Overview of the structural models and MACE-based simulation workflow. (a) Representative 3C, 2H, and 9R SiC polytypes. 3C-SiC is shown along the cubic stacking direction, while 2H- and 9R-SiC are shown as Si--C bilayers stacked along the hexagonal/rhombohedral stacking direction, corresponding to the $c$ axis. (b) Fine-tuned MACE model used for phase-coexistence melting simulations and finite-temperature phonon analysis. The final dataset size and model errors are reported.}
    \label{fig:mace_DFPT}
\end{figure*}

Figure~\ref{fig:mace_DFPT} summarizes the representative SiC polytypes and the computational workflow used to investigate polytype-dependent melting. We first fine-tuned and validated a MACE interatomic potential for SiC using DFT data generated from crystalline, thermally distorted, and high-temperature molecular-dynamics configurations (Methods). The resulting model reproduces reference DFT energies, forces, and stresses with low errors across crystalline and disordered configurations (energy MAE = 1.20~meV/atom, force MAE = 0.075~eV/\AA{}, and stress errors below 1.3~meV/\AA{}$^{3}$). It also reproduces the 0~K DFPT phonon dispersions of the representative polytypes along the selected $q$-paths (Fig.~S2). Detailed validation results, including error breakdowns, parity plots, and descriptor analyses, are provided in the Supporting Information (Figs.~S1 and S3, Tables~S1 and S2).

We next examine how structural and lattice-dynamical differences between these polytypes are reflected in high-temperature melting behavior.

\subsection{Microscopic melting mechanism}

We first investigate the microscopic melting mechanism of SiC polytypes using phase-coexistence molecular dynamics simulations. An initial upper-bound estimate of $\sim$3500~K obtained from single-phase heating is reduced when phase coexistence is introduced, consistent with the suppression of superheating. The melting behavior of the 3C, 2H, and 9R polytypes is characterized using radial distribution functions (RDFs), a disorder fraction $f$, and mean-squared displacements (MSDs) (Figs.~\ref{fig:rdf}--\ref{fig:CN_MSD}).

Before melting, all polytypes exhibit sharp Si--C coordination peaks at $r \approx 1.9$~\AA{}, arising from the tetrahedral Si--C network, while Si--Si and C--C peaks appear at larger distances ($r \approx 3.0$~\AA{}) associated with the second-neighbor region (Fig.~\ref{fig:rdf}a). Upon heating, the earliest structural signature of destabilization is the formation of short C--C contacts at $r \approx 1.5$~\AA{}, close to typical C--C bond lengths (1.42~\AA{} in graphite and 1.54~\AA{} in diamond) (Fig.~\ref{fig:rdf}b). This indicates the formation of carbon-rich clusters within the Si--C framework. The Si--C nearest-neighbor peak remains pronounced during the early stages, showing that local bond rearrangements precede the loss of long-range order. With further heating, the C--C peak increases in intensity, the Si--C peak broadens and weakens, and a weak Si--Si peak appears at $r \approx 2.5$~\AA{}, indicating the formation of short Si--Si contacts (Fig.~\ref{fig:rdf}c).

These RDF changes reflect the progressive replacement of Si--C bonds by C--C and Si--Si interactions at high temperatures. To quantify this disordering, we define the disorder fraction as
\(f = [N_{\mathrm{C-C}}(r<r_c)+N_{\mathrm{Si-Si}}(r<r_c)]/N_{\mathrm{tot}}(r<r_c)\),
where \(N_{\mathrm{tot}}\) is the total number of first-shell contacts and \(r_c=2.35~\text{\AA}\). The disorder fraction increases sharply near melting, providing a compact indicator of the transition (Fig.~\ref{fig:CN_MSD}a).

This analysis indicates a two-step melting mechanism: (i) formation of short C--C contacts and carbon-rich regions within the crystalline framework, followed by (ii) progressive loss of Si--C tetrahedral connectivity and long-range order. The same pathway is visible in the coexistence snapshots, which show the development of disordered regions and short C--C contact networks during melting (Fig.~\ref{fig:snapshots}). Short Si--Si contacts appear at later stages as the structure becomes increasingly disordered. This picture is consistent with recent simulations reporting incongruent melting via carbon clustering~\cite{Xie2026}. The short C--C contacts observed in our simulations represent microscopic precursors to melting in all studied polytypes.

\subsection{Polytype-dependent melting in SiC}

\begin{figure}[h!]
    \centering
    \includegraphics[width=0.5\textwidth]{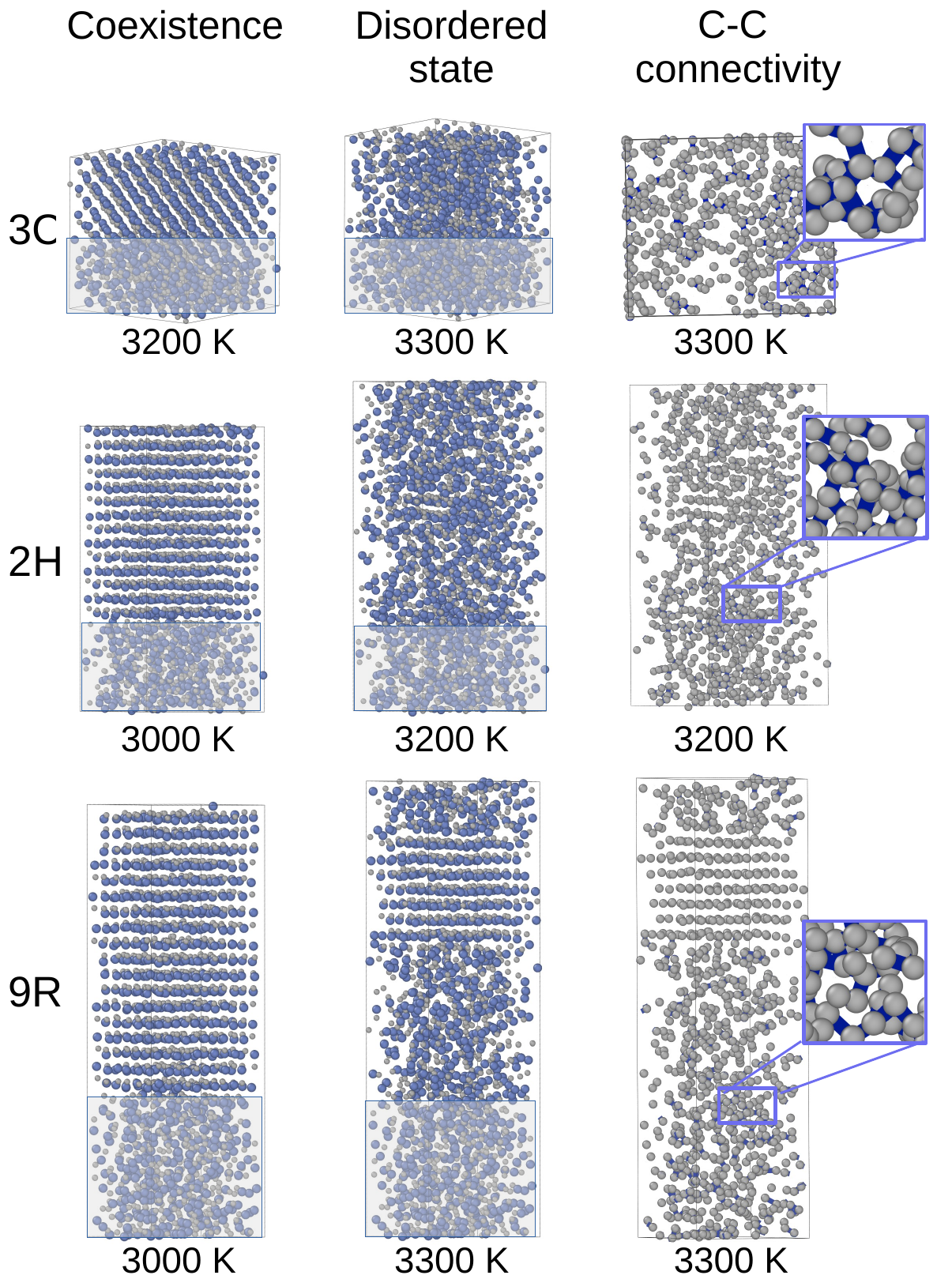}
    \caption{Phase-coexistence melting simulations of 3C, 2H, and 9R SiC. The first column shows representative two-phase solid--liquid configurations near the coexistence regime, while the second column shows high-temperature disordered configurations. The third column shows only short C--C contact regions, with insets showing local C--C connectivity.}
    \label{fig:snapshots}
\end{figure}

\begin{figure*}[h!]
    \centering
    \includegraphics[width=0.95\textwidth]{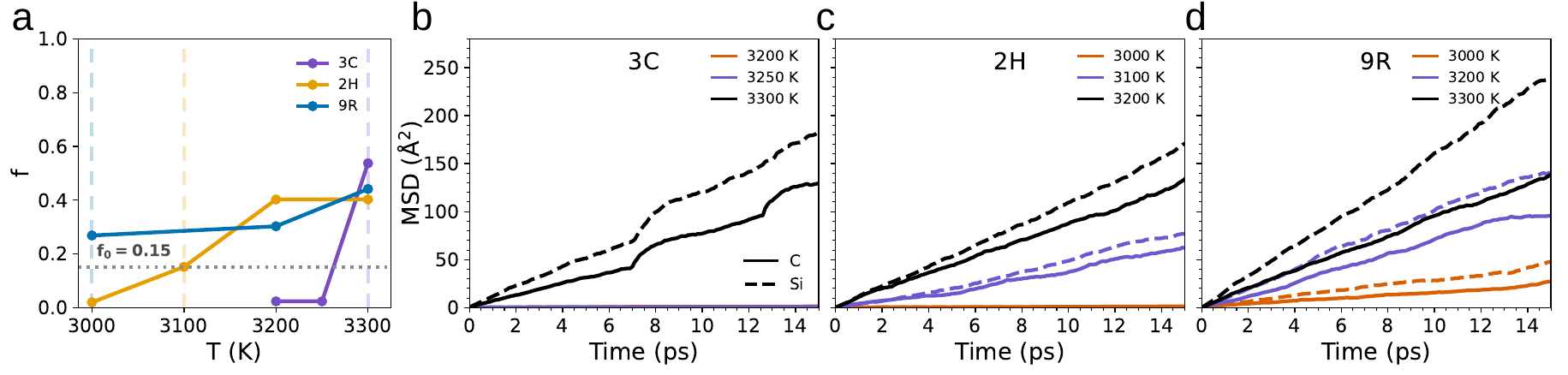}
    \caption{Structural and dynamical indicators of melting from phase-coexistence MD trajectories.(a) Disorder fraction $f$ for 3C, 2H, and 9R SiC, computed from first-neighbor coordination numbers ($r<2.35$~\AA{}) as the fraction of homonuclear contacts. The horizontal dotted line marks melting initiation, and vertical dashed lines indicate the corresponding temperatures. (b--d) MSDs of C (solid lines) and Si (dashed lines) atoms at selected temperatures for 3C, 2H, and 9R SiC, shown over the final 15 ps window after melting is established and before substantial volume expansion or sublimation.} 
    \label{fig:CN_MSD}
\end{figure*}

Melting intervals were estimated as the lowest temperatures at which disorder signatures in RDFs appear and persist during extended coexistence simulations. The disorder fraction threshold \(f_0=0.15\) (Fig.~\ref{fig:CN_MSD}a) is used as a practical indicator of melting initiation. Within the temperature resolution of the simulations, 3C-SiC exhibits the highest thermal stability (\(T_m \approx 3300\)~K), followed by 2H-SiC (\(T_m \approx 3100\)--\(3200\)~K), while 9R-SiC shows melting signatures already at \(\sim 3000\)~K. These trends are consistent with the coexistence snapshots (Fig.~\ref{fig:snapshots}), with 3C maintaining crystalline order up to $3300$~K, whereas 2H and 9R undergo earlier destabilization.

While the RDFs and disorder fraction indicate the structural changes upon heating, they do not distinguish local amorphous-like disorder from a melted, diffusive state. MSDs therefore provide a complementary dynamical criterion for melting (Fig.~\ref{fig:CN_MSD}b--d). Above the melting range, the MSD curves show approximately linear growth indicative of Einstein diffusion. The MSD trends are consistent with the RDF results, confirming that 3C remains stable to the highest temperatures, 2H disorders at intermediate temperatures, and 9R is destabilized first. In all polytypes, Si atoms exhibit larger mobility than carbon in the disordered regime, consistent with the formation of carbon-rich regions. 

\begin{figure*}[h!]
    \centering
    \includegraphics[width=0.7\textwidth]{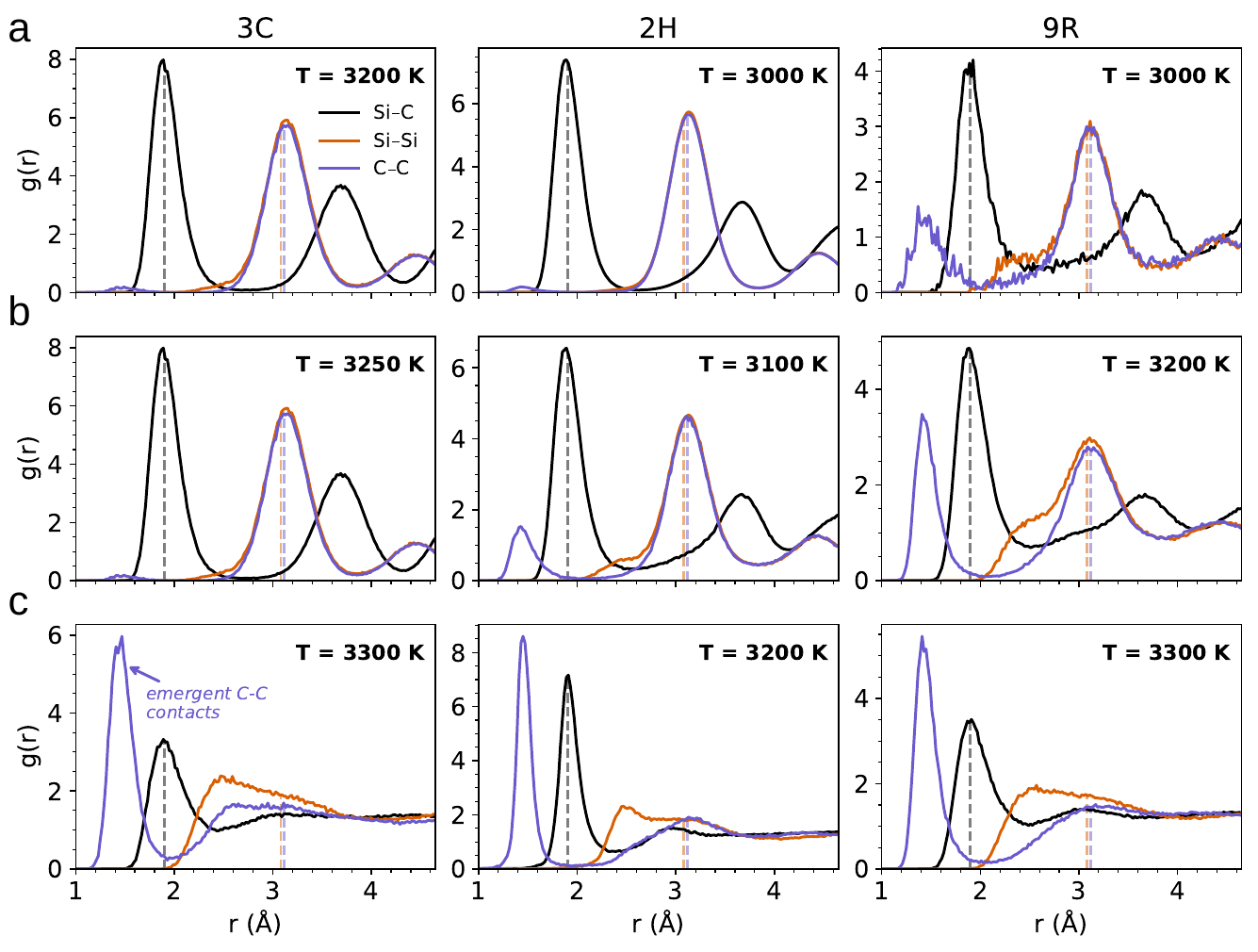}
    \caption{Short- and medium-range radial distribution functions, $g(r)$, for Si--C (black), Si--Si (red), and C--C (blue) pairs in 3C, 2H, and 9R SiC polytypes at selected temperatures. Rows show representative stages of the high-temperature disordering pathway: (a) pre/early-melting stage; (b) increased short C--C contact formation; and (c) advanced disordering/melting, with temperatures chosen separately for each polytype and indicated in each panel. Dashed vertical lines mark 0~K reference distances: nearest-neighbor Si--C ($\sim 1.9$~\AA, black) and second-shell Si--Si/C--C distances ($\sim 3.1$~\AA, red/blue).}
    \label{fig:rdf}
\end{figure*}

In the 9R polytype, local disordering emerges in the nearest-neighbor shell already at \(\sim 3000\)~K, while medium-range order persists to higher temperatures (Figs.~\ref{fig:rdf} and \ref{fig:snapshots}). This indicates a separation between local bond rearrangement and the loss of medium-range order in 9R-SiC. The enhanced mobility observed in 9R at elevated temperatures (\(3300\)~K) further supports its reduced thermal stability.

\subsection{High-temperature phonon dispersion and mode softening in SiC polytypes}
\label{sec:finite_temperature_phonons}

\begin{figure*}[h!]
    \centering
    \includegraphics[width=0.8\textwidth]{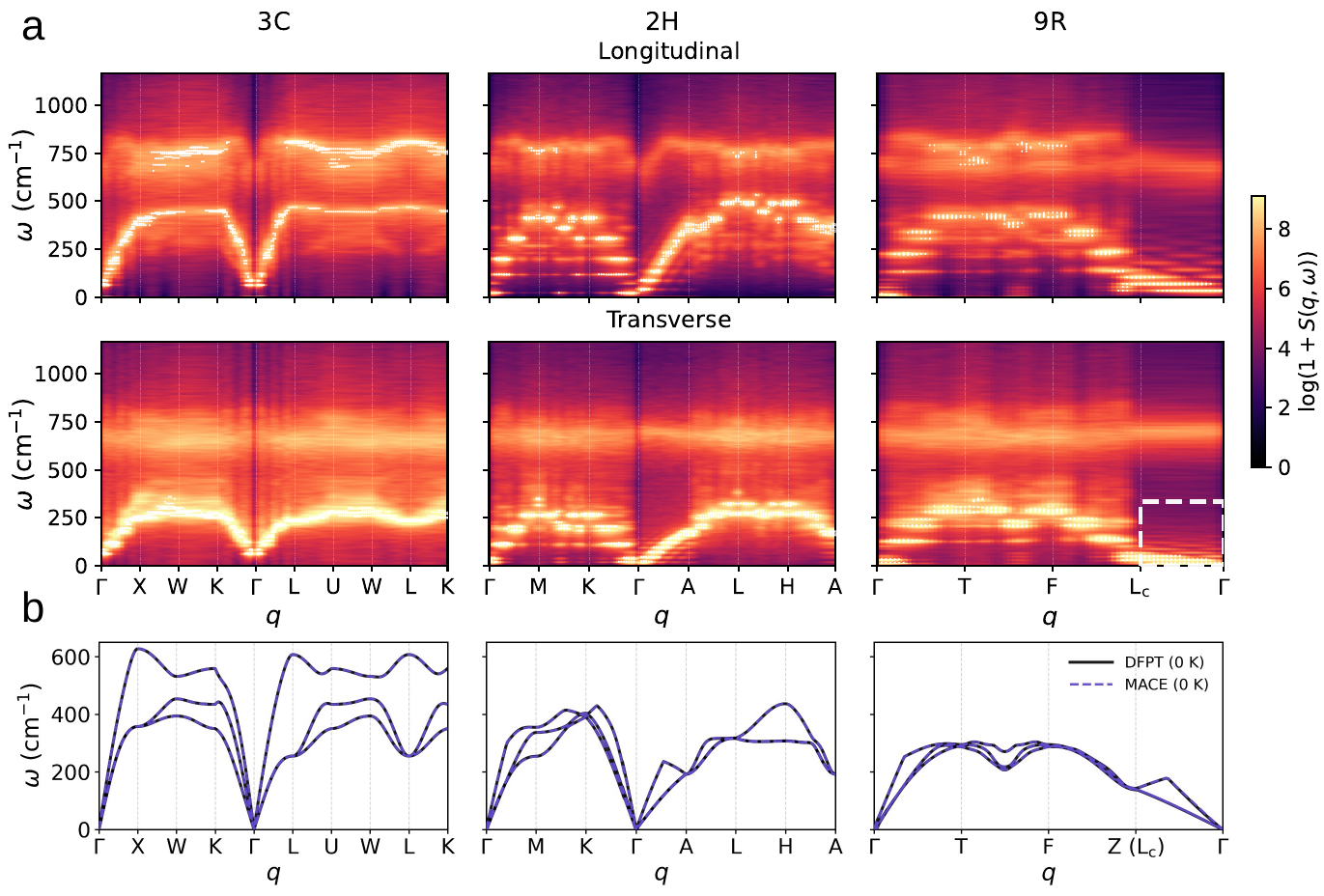}
    \caption{Finite-temperature phonon spectra and 0~K acoustic reference for the 3C, 2H, and 9R SiC polytypes. (a) Longitudinal and transverse components of the current spectra $S(q,\omega)$ at $T=2800$~K, with peak positions overlaid in the brightest color. The dashed rectangle highlights the low-frequency transverse shear-mode region. (b) Acoustic branches at 0~K from DFPT and the fine-tuned MACE model, plotted along the same $q$-paths. For 9R-SiC, the primitive-cell $Z$ point corresponds to the $\mathrm{L}_{\mathrm{c}}$ point used in the conventional cell.}
    \label{fig:phonons}
\end{figure*}

To elucidate the dynamical origin of the polytype-dependent melting behavior, we computed finite-temperature phonon spectra using velocity-current spectral analysis of MACE molecular dynamics trajectories (Fig.~\ref{fig:phonons} and Figs.~S5--S7). The spectra were evaluated at temperatures where the systems remain crystalline. For each polytype, longitudinal and transverse current spectra, denoted here as $S(\mathbf{q},\omega)$, were calculated along the complete high-symmetry paths shown in Fig.~\ref{fig:phonons}. These paths include the stacking-direction segments $\Gamma \rightarrow L$ for 3C, $\Gamma \rightarrow A$ for 2H, and $\Gamma \rightarrow L_c$ for 9R, which are the focus of the shear-mode analysis below. These segments probe modes normal to the close-packed bilayers and therefore capture transverse shear motions associated with relative bilayer sliding. The transverse component represents the combined spectral weight of the two polarizations perpendicular to $\mathbf{q}$.

The main acoustic features in the finite-temperature spectra at 2800~K remain consistent with the 0~K DFPT and MACE acoustic references shown in Fig.~\ref{fig:phonons}b, apart from the expected decrease in frequency due to thermal softening. At 2800~K, longitudinal-acoustic and optical modes remain at high frequencies ($>500$~cm$^{-1}$) and exhibit only moderate thermal broadening in all polytypes, reflecting the strong covalent Si--C bonds within each bilayer. No pronounced softening is observed in these branches over the investigated temperature range.

The transverse-acoustic (TA) modes along the stacking direction show a much stronger polytype dependence. In 3C, the TA branch along $\Gamma \rightarrow L$ remains well defined, with frequencies around $200$--$250$~cm$^{-1}$ and relatively narrow linewidths. The 2H polytype shows similar behavior with slightly lower frequencies and moderate damping. By contrast, 9R exhibits a dense manifold of intrinsically low-frequency TA branches ($\sim 50$--$100$~cm$^{-1}$; Fig.~\ref{fig:phonons}a), associated with transverse shear motion and relative lateral sliding of adjacent bilayers. The long-period stacking sequence also increases the density of low-frequency branches. Several of these branches broaden or become less well resolved at elevated temperatures, consistent with stronger anharmonic dynamics. Similar trends are observed at temperatures between 2000 and 3000~K (Fig.~S5--S7).

In the classical high-temperature regime, each phonon mode carries an average energy of approximately $k_{\mathrm{B}}T$, and the mean-square atomic displacement scales as $\langle u^{2} \rangle \propto k_{\mathrm{B}}T/\omega^{2}$. The low-frequency TA shear modes of 9R are therefore expected to generate larger thermal lateral displacements than the higher-frequency shear modes of 3C and 2H. At finite temperature, this difference is evident from the zone-boundary TA frequencies along the stacking direction, evaluated at the $L$, $A$, and $L_c$ points for 3C, 2H, and 9R, respectively. While the 3C and 2H modes remain in the range of approximately 150--250~cm$^{-1}$, the corresponding 9R mode remains below 50~cm$^{-1}$ over the analyzed crystalline temperature range before melting (2000--3000~K). Interestingly, the same frequency ordering is already present in the 0~K acoustic branches shown in Fig.~\ref{fig:phonons}b. The zone-boundary TA mode along the stacking direction has the lowest frequency in 9R-SiC, with $\omega_{\mathrm{TA}} \approx 125$~cm$^{-1}$ at the $L_c/Z$ point, compared with $\approx 190$~cm$^{-1}$ at $A$ in 2H-SiC and $\approx 254$~cm$^{-1}$ at $L$ in 3C-SiC. At finite temperature, this initially lower mode softens more strongly in 9R than in 2H or 3C.

\begin{figure*}[h!]
    \centering
    \includegraphics[width=0.7\textwidth]{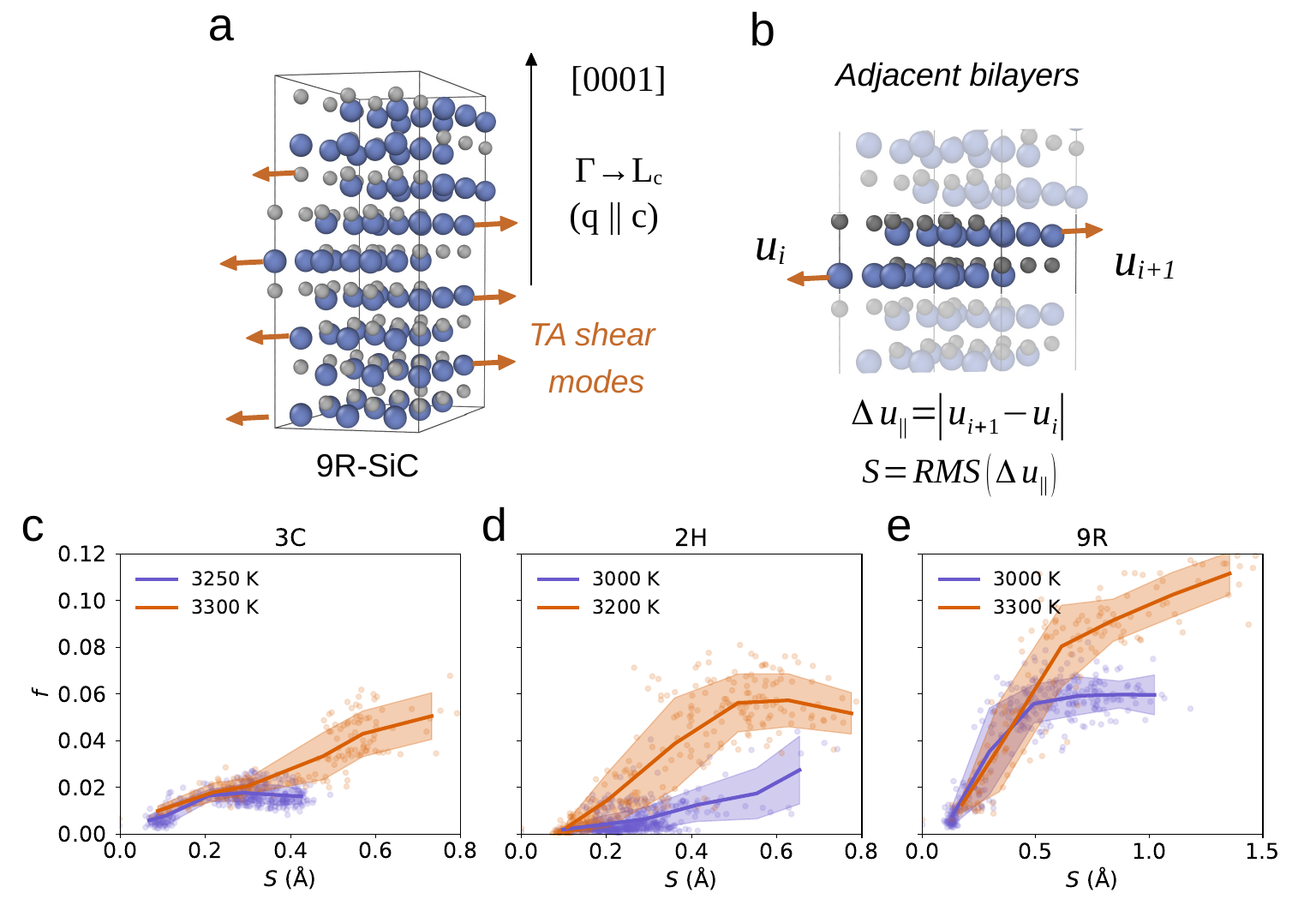}
\caption{Stacking-direction shear modes and disordering in SiC polytypes. (a) Transverse-acoustic shear displacements in 9R-SiC along [0001], illustrating lateral sliding of adjacent bilayers. (b) Definition of the sliding metric $S$. (c)--(e) Disorder fraction $f$ as a function of $S$ for 3C, 2H, and 9R at representative temperatures near melting. Symbols denote MD configurations, lines show binned averages, and shaded regions show the spread within each bin. Fully melted configurations are excluded.}
    \label{fig:9R_modes}
\end{figure*}

To quantify the connection between interlayer sliding and structural disordering, we define a sliding metric $S$ based on the relative in-plane displacements between adjacent close-packed Si--C bilayers (Fig.~\ref{fig:9R_modes}a,b). We then evaluate the correlation between $S$ and the disorder fraction $f$ for representative temperatures near melting (Fig.~\ref{fig:9R_modes}c--e). In all polytypes, larger $S$ is associated with larger $f$, indicating that enhanced lateral sliding accompanies the formation of local structural disorder. The largest sliding amplitudes are observed for 9R within the analyzed trajectory segments, supporting the interpretation that its low-frequency TA shear modes promote bilayer sliding and facilitate the formation of disordered precursors to melting.

The transverse spectrum of 9R therefore differs from those of 3C and 2H mainly through the intrinsically lower frequencies of the stacking-direction TA modes. These low-frequency modes enhance thermal lateral displacements and make the long-period stacking sequence more susceptible to local disordering during bilayer sliding.

\section{Discussion}

Experimentally, SiC melting and decomposition have been reported over a broad temperature range, with onset temperatures depending strongly on pressure, methodology, and the interpretation of congruent versus incongruent melting~\cite{Sokolov2012,cryst8050217}. Our phase-coexistence simulations were performed near ambient pressure, at 1~bar, and predict melting intervals within this broad experimental range, although polytype-resolved experimental data are not available. The formation of spatially distributed carbon-rich local regions is consistent with local, incongruent-like disordering, although our simulations do not establish macroscopic phase separation or the formation of a distinct crystalline carbon phase. Recent machine-learning molecular dynamics studies have addressed SiC decomposition and incongruent melting at substantially higher pressures, providing a complementary high-pressure perspective~\cite{Xie2026}.

We focus here on establishing the high-temperature stability ordering of representative SiC polytypes and its dependence on crystal structure. The local chemical pathway is similar in all studied polytypes: short C--C contacts and carbon-rich regions appear first, followed by progressive loss of tetrahedral Si--C connectivity. The polytype dependence therefore does not arise from a different local precursor, but from how the stacking sequence modifies the lattice dynamics that activate this precursor. In 9R, lower-frequency stacking-direction TA modes enhance thermally activated bilayer sliding, and the positive correlation between the sliding metric \(S\) and disorder fraction \(f\) links this sliding to local chemical disordering. This provides a microscopic explanation for the observed stability ordering \(3\mathrm{C} > 2\mathrm{H} > 9\mathrm{R}\), and places the polytype-dependent melting behavior in the context of the Born criterion for melting, which relates melting to the loss of shear stability~\cite{Born1939}. In the present case, this shear-softening picture emerges through finite-temperature dynamics rather than only through static elastic stability. The fact that the relevant TA modes in 9R are already lowest in frequency at 0~K indicates that the reduced shear stability originates from harmonic lattice dynamics and is further enhanced by anharmonic effects at high temperature. This suggests that harmonic stacking-direction TA modes may provide useful proxies for relative high-temperature stability in SiC polytypes.

Direct experimental measurements of melting temperatures for different SiC polytypes are, to our knowledge, not available. However, related evidence from layered and polytypic materials shows that stacking can strongly affect vibrational and energetic properties even when the local bonding environment is similar. In layered materials, low-frequency vibrational modes are strongly affected by interlayer coupling and stacking, including bending modes in graphite/graphene~\cite{Marzari2005} and interlayer shear modes in few-layer graphene and transition-metal dichalcogenides~\cite{Luo2015}. Although SiC is not a van der Waals layered crystal, the stacking-dependent effects may still influence vibrational and thermodynamic behavior. Related stacking-dependent structural and energetic trends have also been reported in tetrahedrally coordinated polytypic semiconductors such as ZnS, ZnSe, and ZnTe, where closely related local bonding leads to small polytype energy differences and the 3C polytype is the most stable structure~\cite{Bechstedt2014}. In this context, our results suggest that identical local bonding but different medium- and long-range stacking order can lead to measurable differences in vibrational, structural, and high-temperature properties.  The more isotropic 3C network retains greater resistance to shear-like distortions, whereas the hexagonal/rhombohedral bilayer stacking sequences in 2H and especially 9R make relative bilayer sliding more accessible, reducing the high-temperature stability of these polytypes.

The fine-tuned MACE model has sufficient structural sensitivity to resolve these coupled structural and dynamical effects while enabling the length and time scales required to observe crystalline--disordered transformations beyond the reach of direct first-principles simulations.

\section{Conclusions}

In this work, we combined phase-coexistence machine-learning molecular dynamics with finite-temperature phonon analysis to resolve the microscopic origin of polytype-dependent melting in SiC. A fine-tuned MACE model was developed to describe crystalline, high-temperature, and disordered configurations across multiple SiC polytypes, enabling simulations of crystalline--disordered transformations at length and time scales inaccessible to first-principles methods.

Our phase-coexistence simulations predict melting intervals that fall within the range of available experimental values for SiC. Across all studied polytypes, melting is initiated by the formation of short C--C contacts and carbon-rich local regions, followed by progressive loss of tetrahedral Si--C connectivity. However, the temperature range over which melting occurs depends strongly on the polytype and stacking sequence. The resulting high-temperature stability ordering, \(3\mathrm{C} > 2\mathrm{H} > 9\mathrm{R}\), is reflected consistently in RDFs, disorder fractions, MSDs, and finite-temperature phonon spectra.

The reduced stability of 9R is traced to its low-frequency transverse-acoustic shear modes along the stacking direction. These modes already have the lowest frequencies in the 0~K harmonic spectrum and soften further at high temperature compared with 3C and 2H, indicating that the reduced stability against shear motion originates from the underlying lattice dynamics and is enhanced by anharmonic effects. Since thermal displacement amplitudes scale approximately as \(\langle u^2\rangle \propto k_{\mathrm{B}}T/\omega^2\), these modes generate larger lateral bilayer displacements. The positive correlation between the sliding metric \(S\) and the disorder fraction \(f\) directly links this enhanced sliding to local chemical disordering and short C--C contact formation. In this sense, our simulations provide atomistic finite-temperature support for the Born criterion for melting: low-frequency TA modes promote bilayer sliding, which facilitates local chemical disordering, short C--C contact formation, and ultimately melting. The fact that the same frequency ordering is already present at 0~K suggests that harmonic stacking-direction TA modes can serve as useful proxies for relative high-temperature stability in SiC polytypes.

More broadly, this work shows that high-temperature stability in polytypic covalent materials is not determined only by local bond strength, but also by stacking-dependent lattice dynamics. The more isotropic 3C network retains higher-frequency shear modes and the highest thermal stability, while the hexagonal/rhombohedral stacking sequences in 2H and especially 9R make relative bilayer sliding more easily activated along the stacking direction. The resulting connection between low-frequency shear modes, sliding, and local disordering highlights stacking-dependent transverse dynamics as a key microscopic factor in the high-temperature stability of polytypic materials.

\section{Computational details}

\subsection{DFT computations}

Density functional theory (DFT) reference data, comprising total energies, forces, and stresses, were generated using the plane-wave pseudopotential method as implemented in Quantum ESPRESSO \cite{Giannozzi2009,Giannozzi2017,Giannozzi2020}. The exchange–correlation was described within the generalized gradient approximation using the PBEsol functional \cite{PBEsol}, which was tested for numerical convergence and compared against two other functionals as well as experimental data (see supporting information). Core–valence interactions were treated using PSLibrary v1.0.0 pseudopotentials: an ultrasoft RRKJ pseudopotential for Si and a PAW dataset for C \cite{DALCORSO2014337}.

The wavefunction and charge-density cutoffs were set to 60 and 600~Ry, respectively. Brillouin-zone sampling used a 12$\times$12$\times$12 Monkhorst--Pack grid for primitive 3C-SiC, while meshes for other polytypes and larger supercells were scaled to maintain a comparable reciprocal-space sampling density~\cite{JPSJ2012}. Electronic occupations were treated using Marzari--Vanderbilt smearing~\cite{Marzari1999} with a width of 0.005~Ry, and self-consistency was converged to $10^{-7}$~Ry. Forces and stresses were computed in all calculations. Harmonic phonon dispersions at 0~K were also calculated for optimized primitive cells using DFPT as implemented in Quantum ESPRESSO and finite-displacements with the fine-tuned MACE model as implemented in \texttt{phonopy}~\cite{phonopy-phono3py-JPCM,phonopy-phono3py-JPSJ}. Full details and comparisons are provided in the Supporting Information.

\subsection{MACE model fine-tuning}

All melting simulations were performed using a fine-tuned MACE model obtained by fine-tuning the pretrained MACE-MP-0b2 foundation model~\cite{batatia2023foundation}. Preliminary tests across available MACE-MP variants showed comparable accuracy for 3C-SiC; MACE-MP-0b2 was selected because its smaller cutoff radius of 5~\AA{} enables more efficient MD simulations, while the message-passing architecture extends the interaction range to 10~\AA{}.

The initial training dataset was constructed from strained and rattled crystalline configurations~\cite{magorrian2025}, but this dataset was insufficient to describe bond-breaking and strongly anharmonic environments. It was therefore extended iteratively with configurations sampled from MD trajectories. Initial 3C-SiC simulations were performed at 1~bar up to 4500~K using the original MACE-MP-0b2 model until melting occurred. Representative crystalline, thermally perturbed, defected, and partially disordered configurations, including structures with short C--C and Si--Si contacts, were recomputed at the DFT level to obtain reference energies, forces, and stresses and then added to the training set.

This iterative cycle of MD sampling, DFT recomputation, and model retraining was first applied to 3C and subsequently extended to 2H, 4H, 6H, 8H, and 9R, ensuring consistent coverage of configurational space across polytypes. The final dataset comprised 12,661 configurations, including 10,065 training, 1,468 validation, and 1,128 test structures; a detailed breakdown is provided in Table~S1. Configurations were labeled as disordered when they contained at least one short homonuclear contact, C--C or Si--Si, below 1.7~\AA{}. Since such contacts are absent in crystalline SiC, this criterion identifies configurations involving bond rearrangements and clustering rather than purely thermal broadening.

Training was initialized from the weights of the MACE-MP-0b2 model, with all parameters allowed to vary during fine-tuning. The model employs 128 invariant and 128 equivariant feature channels across two interaction layers with symmetry order $L=1$ and correlation order $\nu=3$. Atomic environments were represented using a cutoff radius of $5\,\text{\AA}$. Fine-tuning was performed using a batch size of 10. The relative weights of the energy, force, and stress contributions to the loss function were set to $1:100:1$. Optimization employed the AMSGrad variant of the Adam optimizer~\cite{Kingma2014AdamAM, Adam2018}, with an initial learning rate of $10^{-2}$, reduced to $10^{-4}$ after 200 epochs to stabilize convergence.

\subsection{Phase-Coexistence Melting Simulations}

The final fine-tuned model was used for all subsequent MD simulations under periodic boundary conditions. Melting temperatures were estimated using the phase-coexistence approach~\cite{Car1995,Alfe2002}, which suppresses superheating by introducing explicit solid--liquid interfaces.

\paragraph{Single-phase heating.}

Prismatic supercells of the 3C, 2H, and 9R polytypes were constructed and elongated along the stacking ($z$) direction to enable consistent solid--liquid interface construction. The total numbers of atoms were 1728, 1944, and 1960 for 3C, 2H, and 9R, respectively. These cell sizes provide comparable cell shapes across polytypes while remaining computationally feasible for near-melting coexistence simulations and finite-temperature vibrational analysis. They are consistent with recent MACE-based studies of phase transitions, melting, and vibrational dynamics using systems of similar size~\cite{Coudert2026,DellaPia2024,Rossi2025}. Each system was equilibrated at 300~K and then heated to $T_i = 3000~\text{K}$ at a rate of $h_r = 10~\text{K/ps}$ in the NPT ensemble at 1~bar, generating equilibrated high-temperature crystalline structures.

\paragraph{Construction of two-phase configurations.}

Each supercell was divided along the stacking direction into a larger crystalline region ($\approx$ $2/3$ of the cell) and a smaller region ($\approx$ $1/3$) designated for melting. The crystalline region was temporarily constrained and maintained at $T_i$, while the unconstrained region was gradually heated from $T_i$ to temperatures in the range $3000~\text{K} \leq T \leq 4200~\text{K}$ in the anisotropic NP$_z$T ensemble. The in-plane cell dimensions were fixed, while the $z$ dimension was controlled by a barostat at $p_z = 1~\text{bar}$. Once a stable disordered region formed, it was frozen to provide a nucleation centre.

\paragraph{Estimation of melting temperatures.}

With the disordered region fixed, the crystalline region was heated from $3000$ to $4200~\text{K}$ at a reduced rate of $h_r = 1~\text{K/ps}$ in the NP$_z$T ensemble, which allows relaxation normal to the interface while keeping the lateral cell dimensions fixed~\cite{Espinosa2013}. The temperature at which the crystalline portion irreversibly melted provided an upper-bound estimate, $T_m^{\mathrm{max}}$. This estimate was refined by performing coexistence simulations at progressively lower temperatures ($T_m^{\mathrm{max}} - 100~\text{K}$, $T_m^{\mathrm{max}} - 200~\text{K}$, \ldots) until the lowest temperature at which persistent disorder was observed was identified. The melting temperature was therefore assigned within the interval between the highest temperature without melting and the lowest temperature with irreversible melting. Each coexistence simulation was run for up to 0.5~ns or until complete melting.

Temperature and pressure were controlled using a Nos\'e–Hoover thermostat and barostat~\cite{Hoover1985,NHthermostat,Martyna1994,Martyna10041996} with damping parameters of 0.5 ps and 5.0 ps, respectively. A time step of 1~fs was used. All simulations were performed using GPU-accelerated LAMMPS~\cite{lammps}.

\subsection{Finite-Temperature Phonon Dispersions}

Finite-temperature phonon spectra were obtained from a $q$-resolved velocity-current spectral analysis of the MD trajectories using an in-house implementation. The analysis follows the current-correlation formalism, in which longitudinal and transverse current spectra are constructed directly from atomic velocities and positions~\cite{Hove1954,HansenMcDonald2013,Fransson2021}. Temperature-renormalized phonon peak positions were then identified from the maxima of the resulting spectra along the selected $q$-paths. This current-based analysis does not require projection onto harmonic eigenmodes.

For each polytype the system was equilibrated at four temperatures below the melting point (2000~K, 2400~K, 2800~K, and 3000~K), followed by 40--80~ps of production MD. Atomic positions and velocities were recorded every 10~fs for subsequent analysis.

For each wavevector $\mathbf{q}$ along the chosen high-symmetry paths, we evaluated the microscopic velocity-current

\begin{equation}
\mathbf{J}(\mathbf{q},t) = 
\frac{1}{N}
\sum_{j=1}^{N} 
\mathbf{v}_j(t)\,
e^{-i \mathbf{q}\cdot \mathbf{r}_j(t)},
\end{equation}
where $\mathbf{r}_j(t)$ and $\mathbf{v}_j(t)$ denote the position and velocity of atom $j$ at time $t$. An orthonormal basis $\{\hat{\mathbf{q}},\mathbf{e}_1,\mathbf{e}_2\}$ was constructed with $\hat{\mathbf{q}}=\mathbf{q}/|\mathbf{q}|$, allowing decomposition into longitudinal and transverse components,
\begin{equation}
J_L(\mathbf{q},t)=\hat{\mathbf{q}}\cdot\mathbf{J}(\mathbf{q},t), 
\qquad
J_T(\mathbf{q},t)=\mathbf{e}_1\cdot\mathbf{J}(\mathbf{q},t) 
+ i\,\mathbf{e}_2\cdot\mathbf{J}(\mathbf{q},t).
\end{equation}

Spectral intensities were obtained from the squared magnitude of the discrete Fourier transform of $J_L$ and $J_T$, evaluated over finite trajectory windows and averaged to improve statistical convergence,
\begin{equation}
S_{L/T}(\mathbf{q},\omega) 
\propto 
\left|
\mathcal{F}
\left[
J_{L/T}(\mathbf{q},t)
\right]
\right|^2.
\end{equation}

The transverse component was constructed from a complex combination of two orthogonal directions perpendicular to $\mathbf{q}$, such that the resulting transverse current spectrum represents the combined intensity associated with motions transverse to $\mathbf{q}$. Renormalized phonon peak positions $\omega_0(\mathbf{q},T)$ were identified from the maxima of $S_{L/T}(\mathbf{q},\omega)$ along each high-symmetry path. This MD-based spectral analysis captures temperature-induced frequency renormalization and phonon damping effects in the strong anharmonic regime close to melting.

\section{Acknowledgments}

This work was supported by the Hartree National Centre for Digital Innovation, a collaboration between STFC and IBM. This work made use of computing resources provided by the STFC Hartree Centre as well as the STFC Scientific Computing Department’s SCARF cluster.

\section{Data Availability}

The DFT training, validation, and test datasets, the fine-tuned MACE model, and supporting data generated in this study are available in the STFC eData repository at \url{https://doi.org/10.5286/edata/973}.

\section{Author Contributions}

L.S. developed the main scientific framing; generated the DFT training dataset from MD simulations; performed model fine-tuning following consultation with F.L.T.; carried out the MD simulations; developed and applied the finite-temperature phonon analysis; performed structural analysis, data interpretation, and visualization; and wrote and revised the manuscript. S.J.M. and L.S. performed the 0 K DFPT phonon calculations. S.J.M. generated the initial set of 0 K DFT training data, and contributed to training early iterations of the MACE model. L.K. performed the DFT convergence testing. R.N.W. provided industrial insight. V.Z. conceived the idea of creating a MACE MLIP for melting simulations of SiC crystals, contributed to 0 K DFT data generation and reviewed the manuscript. All authors agreed with the final version of the manuscript.

\clearpage
\onecolumngrid

\setcounter{page}{1}
\renewcommand{\thepage}{S\arabic{page}}

\setcounter{section}{0}
\setcounter{figure}{0}
\setcounter{table}{0}
\setcounter{equation}{0}

\renewcommand{\thesection}{S\arabic{section}}
\renewcommand{\thefigure}{S\arabic{figure}}
\renewcommand{\thetable}{S\arabic{table}}
\renewcommand{\theequation}{S\arabic{equation}}

\clearpage

\begin{center}
{\Large \textbf{Supporting Information: Microscopic origin of polytype-dependent melting in SiC revealed by machine-learning molecular dynamics}}

\vspace{1em}

Ljiljana Stojanovi\'c, Samuel J. Magorrian, Lara Kabalan, Richard N. White, Fabian L. Thiemann, and Viktor Z\'olyomi

\vspace{0.5em}

Hartree Centre, STFC Daresbury Laboratory, Warrington WA4 4AD, United Kingdom

Lucideon Limited, Stone Business Park, Brooms Road, Stone, Staffordshire, ST15 0SH, UK

IBM Research Europe, Daresbury, WA4 4AD, UK

\vspace{0.5em}

\texttt{ljiljana.stojanovic@stfc.ac.uk}
\end{center}

\renewcommand{\thesection}{S\arabic{section}}
\renewcommand{\thetable}{S\arabic{table}}
\renewcommand{\thefigure}{S\arabic{figure}}

\section{MACE-MP-0b2 finetuning}

\begin{figure*}[h!]
    \centering
    \includegraphics[width=0.8\textwidth]{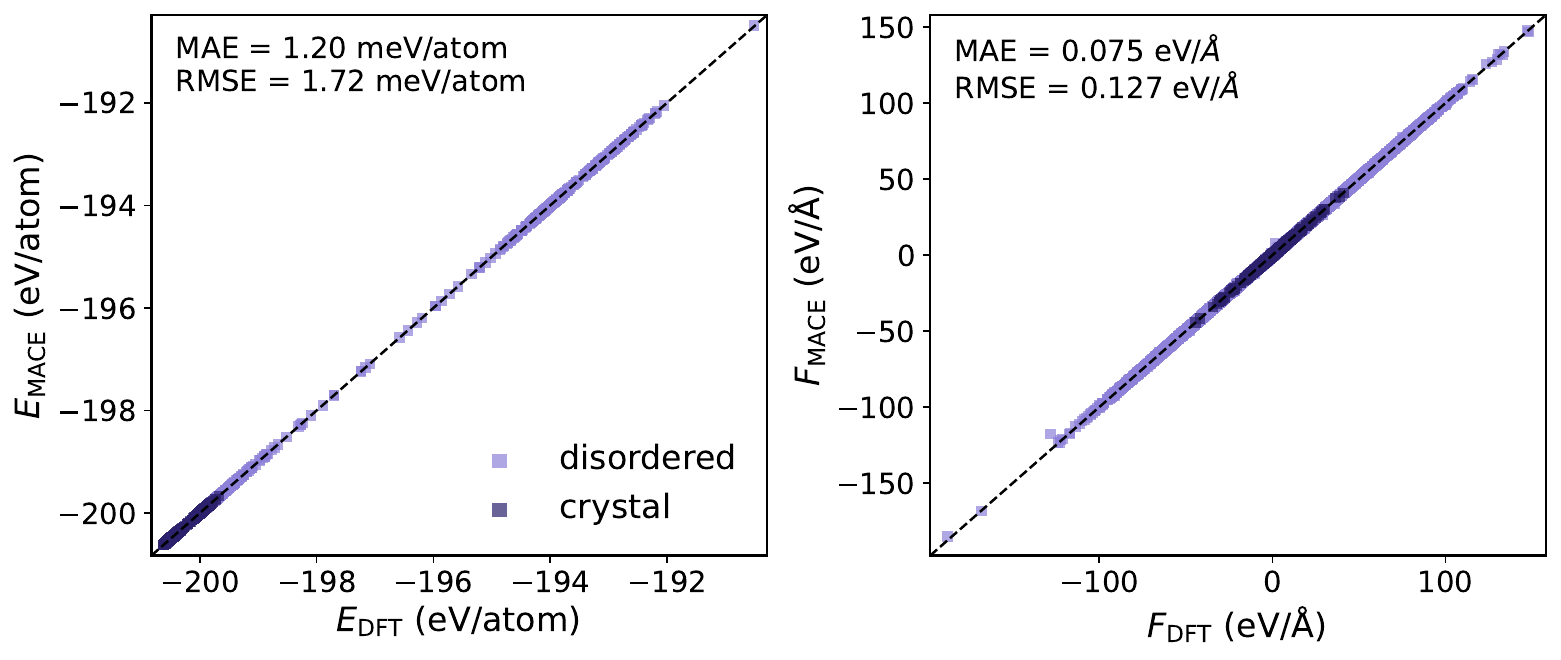}
    \caption{Parity plots comparing MACE predictions and DFT reference values for (a) energies per atom and (b) atomic forces. Data are resolved into crystalline and disordered configurations. The dashed line indicates perfect agreement.}
    \label{fig:parity_plots}
\end{figure*}

Figure~\ref{fig:parity_plots} shows parity plots for energies and forces over the test dataset, including both crystalline and disordered structures. The retrained potential achieves an energy MAE of 1.20 meV/atom (RMSE = 1.72 meV/atom), with no systematic deviation between ordered and disordered environments. Force errors remain low across the entire force range, with an MAE of 0.075 eV/\AA{} and RMSE of 0.127 eV/\AA{}.

A detailed breakdown of energy, force, and stress tensor errors by polytype is provided in Table \ref{tab:mace_errors_by_polytype}. Relative force errors remain close to 1\% across all polytypes, confirming that the model preserves both force directions and magnitudes even in the vicinity of structural instability. Stress (virial) errors remain below 1.3 meV/\AA{}$^{3}$, demonstrating accurate reproduction of elastic and thermodynamic responses. While absolute force errors increase in strongly disordered configurations, this trend correlates with the substantially larger force magnitudes sampled near bond-breaking and highly anharmonic states. Training and validation errors are statistically consistent, indicating no evidence of overfitting. The resulting potential therefore provides a unified and transferable description spanning low-temperature crystalline states and the highly anharmonic configurations encountered near melting.

\begin{table}[h]
\centering
\caption{Training set composition.}
\label{tab:dataset}
\begin{tabular}{lccc}
\hline
Polytype & Crystalline & Disordered & Total \\
\hline
3C  & 2863 & 1308 & 4171 \\
2H  & 1436 & 851  & 2287 \\
4H  & 76   & 476  & 552  \\
6H  & 140  & 218  & 358  \\
8H  & 444  & 526  & 970  \\
9R  & 1407 & 320  & 1727 \\
\hline
Total & 6366 & 3699 & 12661 \\
\hline
\end{tabular}
\end{table}

\begin{table}[h!]
\centering
\caption{Root mean square errors (RMSE) of energies, forces, and stress (virials) for the trained MACE model evaluated on the training, validation, and test datasets, resolved by crystal polytype and disorder.}
\label{tab:mace_errors_by_polytype}
\resizebox{\textwidth}{!}{%
\begin{tabular}{lcccc}
\hline
Config type & 
RMSE $E$ (meV/atom) & 
RMSE $F$ (meV/\AA) & 
Relative $F$ RMSE (\%) & 
RMSE stress (virials) \\
 & & & & (meV/\AA\,\AA$^3$) \\
\hline
Training set    & 1.7 & 109.6 & 0.93 & 0.6 \\
Validation set  & 2.0 & 122.8 & 1.06 & 1.0 \\
\hline
\multicolumn{5}{l}{Test set} \\
2H\_crystal     & 0.5 &  33.0 & 1.53 & 0.4 \\
2H\_disordered  & 3.3 & 198.4 & 1.12 & 1.2 \\
3C\_crystal     & 1.1 &  42.6 & 1.72 & 0.4 \\
3C\_disordered  & 2.4 & 165.6 & 0.91 & 0.9 \\
4H\_crystal     & 1.3 &  61.7 & 2.17 & 0.5 \\
4H\_disordered  & 1.5 & 124.5 & 1.17 & 0.8 \\
6H\_crystal     & 1.2 &  56.1 & 1.99 & 0.5 \\
6H\_disordered  & 1.8 & 148.5 & 1.07 & 0.9 \\
8H\_crystal     & 0.7 &  40.3 & 1.69 & 0.3 \\
8H\_disordered  & 1.6 & 136.2 & 1.08 & 0.9 \\
9R\_crystal     & 1.0 &  47.5 & 1.91 & 0.4 \\
9R\_disordered  & 1.8 & 170.8 & 1.03 & 0.9 \\
\hline
\end{tabular}
}
\end{table}

\clearpage

\subsection{Phonon validation against DFPT at 0~K}

To validate the vibrational properties predicted by the fine-tuned MACE model, we compared 0~K phonon dispersions obtained by density-functional perturbation theory (DFPT) with those calculated with the MACE potential. The DFPT calculations were performed on the optimized primitive cells with the PBEsol functional, pseudopotentials, plane-wave cutoffs, and $k$-point grid as used for the DFT data. Dynamical matrices were computed on uniform $q$-point meshes, and phonon dispersions were evaluated along the selected high-symmetry paths for each polytype. For the primitive rhombohedral 9R cell, the $\Gamma$--Z line, with $Z=(1/2,1/2,1/2)$, corresponds to the stacking direction in the primitive reciprocal basis and is equivalent to the $\Gamma$--$L_c$ direction used in the conventional cell.

The MACE phonon dispersions were also computed on optimized primitive cells using the finite displacement method as implemented in \texttt{phonopy}\cite{phonopy-phono3py-JPCM,phonopy-phono3py-JPSJ}. The same high-symmetry paths were used as in DFPT computations. The comparison shows very good agreement between MACE and DFPT across the acoustic and optical branches for all three polytypes, confirming that the fine-tuned model accurately reproduces phonon spectra of studied polytypes in the harmonic regime. The low-frequency transverse branches associated with shear TA modes are also well captured, supporting the use of MACE for the finite-temperature phonon dispersion.

\begin{figure*}[h!]
    \centering
    \includegraphics[width=0.95\textwidth]{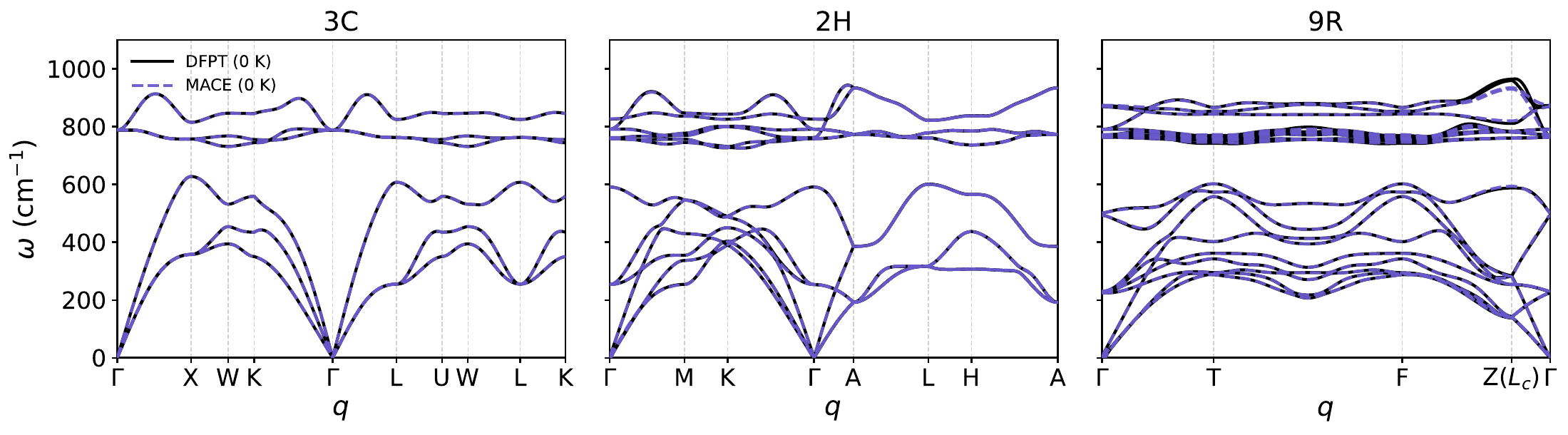}
    \caption{Comparison of 0~K phonon dispersions calculated using DFPT (solid black lines) and the fine-tuned MACE model (dashed blue lines) for 3C, 2H, and 9R primitive cells. Compact high-symmetry paths are used for each polytype: $\Gamma$--X--$\Gamma$--L for 3C, $\Gamma$--M--$\Gamma$--A for 2H, and $\Gamma$--T--$\Gamma$--Z for 9R. }
    \label{fig:0K_phonons}
\end{figure*}

\subsection{Two-dimensional projections of second-layer MACE embeddings}

To characterize the structural diversity captured by the model, we analyze the atomic embeddings generated by the fine-tuned MACE model. Atomic embeddings were extracted from the second interaction layer, averaged over atoms for each configuration, and projected into two dimensions using UMAP~\cite{umap}. Each point represents a configuration corresponding to crystalline or disordered states of the 3C, H-type (2H, 4H, 6H, and 8H), and 9R polytypes, with the projected energy per atom evaluated using the fine-tuned MACE potential.

The embeddings reveal a well-separated crystalline 3C-SiC group, consistent with its cubic stacking. In contrast, the H-type and 9R polytypes partially overlap in descriptor space. As all polytypes share ideal tetrahedral coordination in the first coordination shell, this separation reflects differences beyond the first shell. The 3C polytype exhibits locally isotropic environments, whereas the H-type and 9R structures display a layered character associated with anisotropy along the crystallographic c-axis.

Disordered configurations form a diffuse cloud at higher energies, reflecting the broad configurational space sampled at high temperature. The clear separation between crystalline and disordered regions indicates that the model consistently distinguishes distinct local environments and captures the highly anharmonic configurations relevant for melting.

\begin{figure*}[h!]
    \centering
    \includegraphics[width=0.7\textwidth]{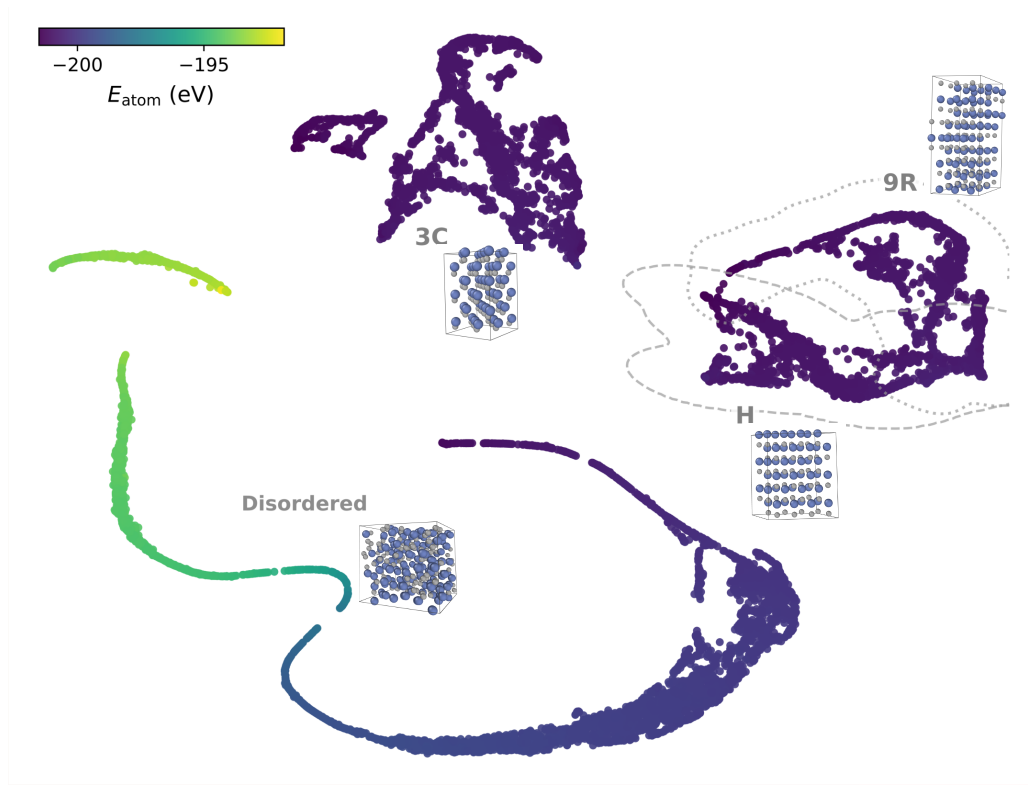}
    \caption{UMAP projections of the SiC embeddings on the dataset used in the model fine-tuning. The crystalline and disordered 3C, hexagonal (H, including 2H, 4H, 6H, and 8H polytypes), and 9R groups are designated, and the energy per atom evaluated using the fine-tuned MACE potential are projected on the structures.}
    \label{fig:mace_umap}
\end{figure*}

\clearpage

\clearpage

\section{Radial distribution functions from phase-coexistence melting simulations}

\begin{figure*}[h!]
    \centering
    \includegraphics[width=0.85\textwidth]{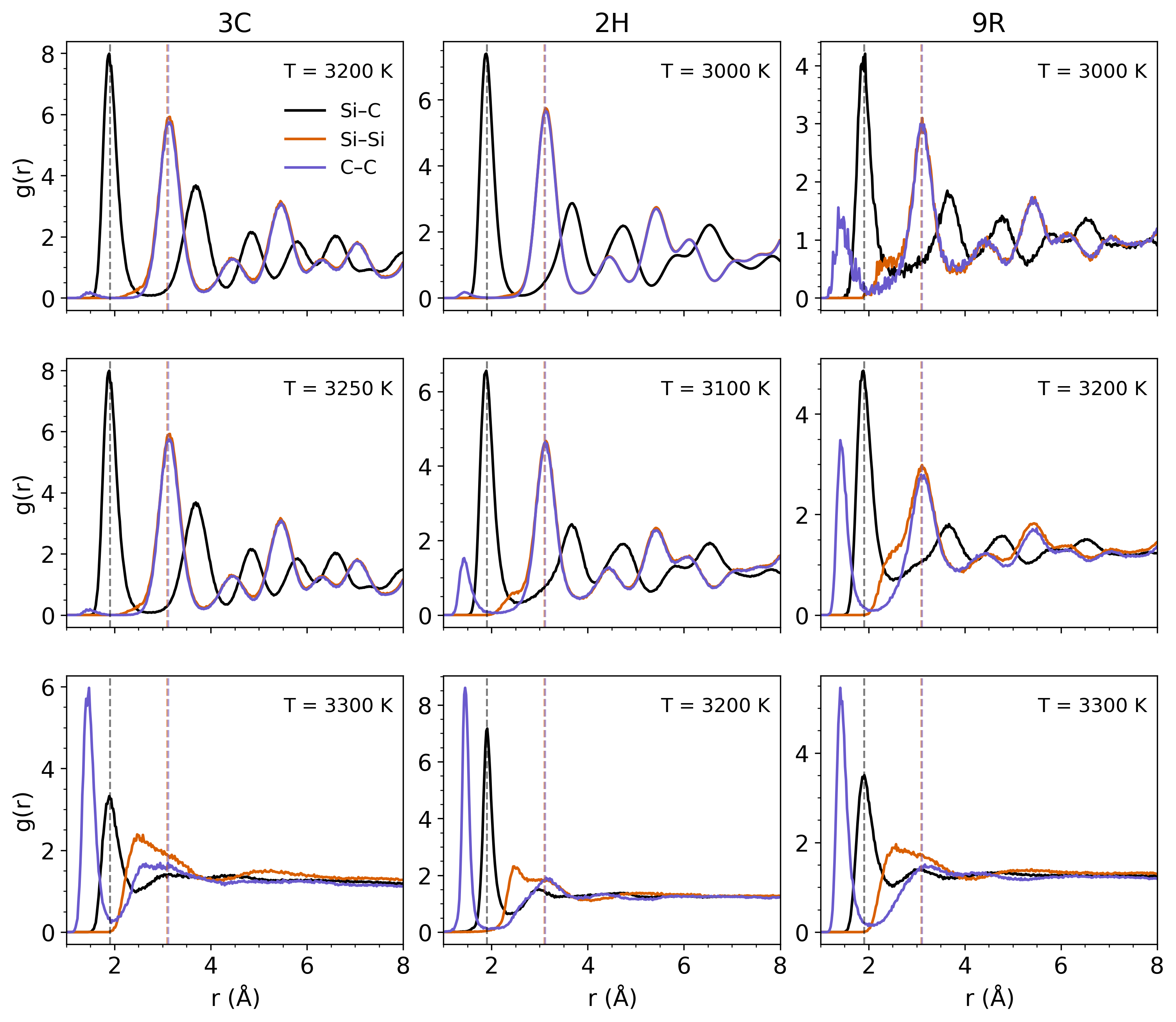}
    \caption{Radial distribution functions ($g(r)$) for Si–C (black), Si–Si (red), and C–C (blue) pairs in 3C, 2H, and 9R SiC polytypes at selected temperatures. Dashed vertical lines indicate reference interatomic distances in the 0 K crystal structure: nearest-neighbor Si–C (~1.9 \AA{}, black) and second-shell Si–Si / C–C  ($\sim 3.1$~\AA, red/blue).}
    \label{fig:rdf_full}
\end{figure*}

Figure \ref{fig:rdf_full} shows the corresponding RDFs over the full radial range. These extended RDFs confirm the same short-range trends discussed in the main text, while also showing the progressive loss of medium- and longer-range order upon heating.

\clearpage

\section{Finite-temperature phonon dispersions}

\subsection{High-symmetry points used for finite-temperature phonon spectra}

\begin{table}[h]
\centering
\caption{High-symmetry points used for the finite-temperature dynamical structure factor calculations. Coordinates are given in fractional reciprocal coordinates of the simulation cell used for each polytype. For 3C and 2H these correspond to the supercells constructed from conventional cubic and hexagonal cells, respectively. For 9R, the finite-temperature simulations were performed using the supercell constructed from conventional 9R cell, for which the point $L_c=(0,0,1/2)$ denotes the zone-boundary point along the stacking ($c$) direction.}
\label{tab:finiteT_qpoints}
\resizebox{\textwidth}{!}{
\begin{tabular}{c c c c}
\hline
Polytype & Point & Fractional coordinate & Role \\
\hline
3C & $\Gamma$ & $(0,0,0)$ & Brillouin zone center \\
3C & $X$ & $(1/2,0,1/2)$ & Cubic/fcc high-symmetry point \\
3C & $W$ & $(1/2,1/4,3/4)$ & Cubic/fcc high-symmetry point \\
3C & $K$ & $(3/8,3/8,3/4)$ & Cubic/fcc high-symmetry point \\
3C & $L$ & $(1/2,1/2,1/2)$ & Boundary along the 3C [111] stacking direction \\
3C & $U$ & $(5/8,1/4,5/8)$ & Cubic/fcc high-symmetry point \\
\hline
2H & $\Gamma$ & $(0,0,0)$ & Brillouin zone center \\
2H & $M$ & $(1/2,0,0)$ & Basal-plane high-symmetry point \\
2H & $K$ & $(1/3,1/3,0)$ & Basal-plane high-symmetry point \\
2H & $A$ & $(0,0,1/2)$ & Boundary along the 2H [0001] stacking direction \\
2H & $L$ & $(1/2,0,1/2)$ & Hexagonal high-symmetry point \\
2H & $H$ & $(1/3,1/3,1/2)$ & Hexagonal high-symmetry point \\
\hline
9R & $\Gamma$ & $(0,0,0)$ & Brillouin zone center \\
9R & $T$ & $(1/2,1/2,0)$ & Conventional cell high-symmetry point \\
9R & $F$ & $(1/2,0,1/2)$ & Conventional cell high-symmetry point \\
9R & $L_c$ & $(0,0,1/2)$ & Boundary along the conventional 9R [0001] stacking direction \\
\hline
\end{tabular}
}
\end{table}

\clearpage

\subsection{Finite-temperature phonon dispersion spectra}

\begin{figure*}[h!]
    \centering
    \includegraphics[width=0.95\textwidth]{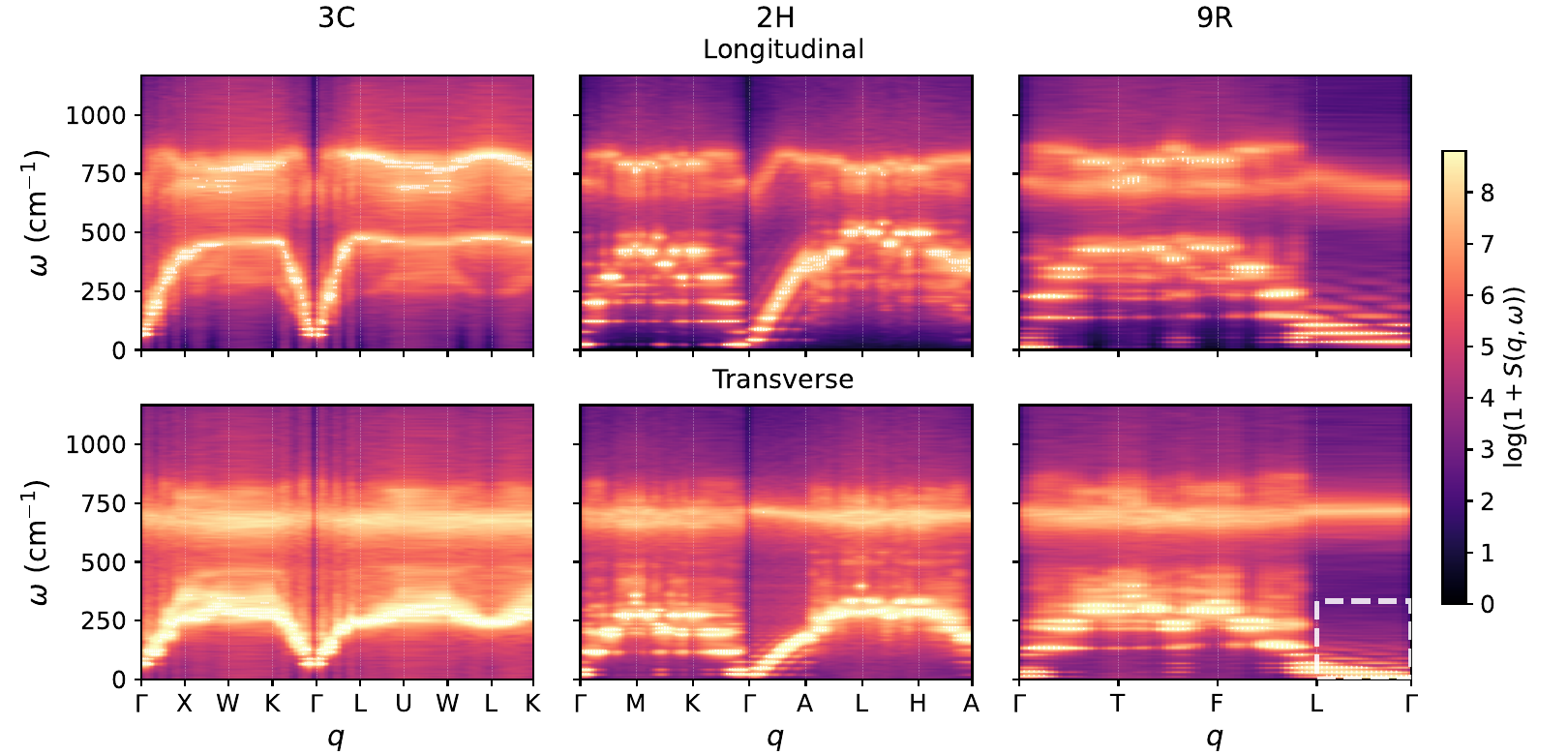}
    \caption{Longitudinal (top row) and transverse (bottom row) components of the dynamical structure factor $S(q,\omega)$ for the 3C, 2H, and 9R SiC polytypes at $T = 2000$~K. White markers denote peak positions extracted from $S(q,\omega)$ maxima. The dashed rectangle highlights the low-frequency shear-mode region in 9R.}
    \label{fig:phonons1}
\end{figure*}

\begin{figure*}[h!]
    \centering
    \includegraphics[width=0.95\textwidth]{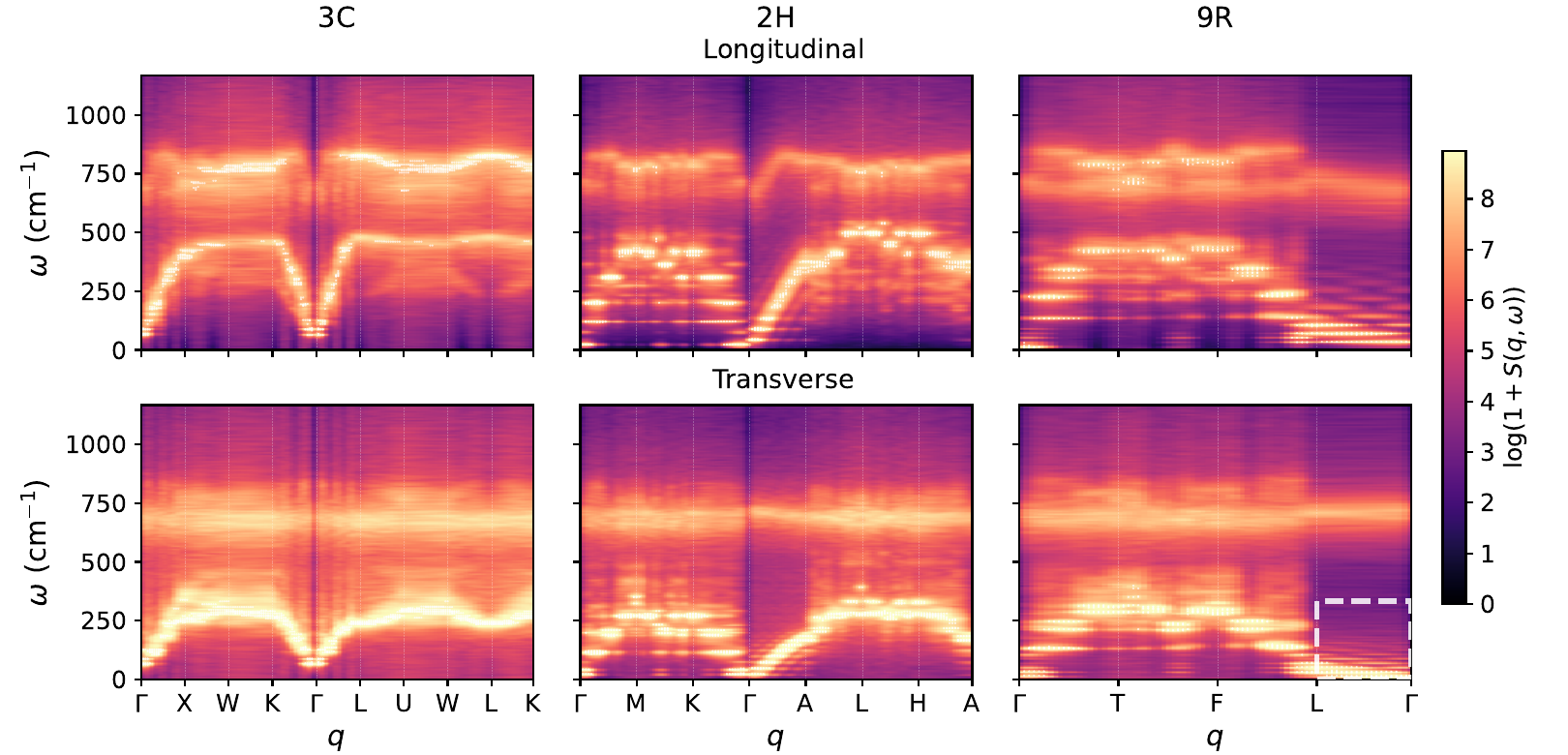}
    \caption{Longitudinal (top row) and transverse (bottom row) components of the dynamical structure factor $S(q,\omega)$ for the 3C, 2H, and 9R SiC polytypes at $T = 2400$~K. White markers denote peak positions extracted from $S(q,\omega)$ maxima. The dashed rectangle highlights the low-frequency shear-mode region in 9R.}
    \label{fig:phonons2}
\end{figure*}

\begin{figure*}[h!]
    \centering
    \includegraphics[width=0.95\textwidth]{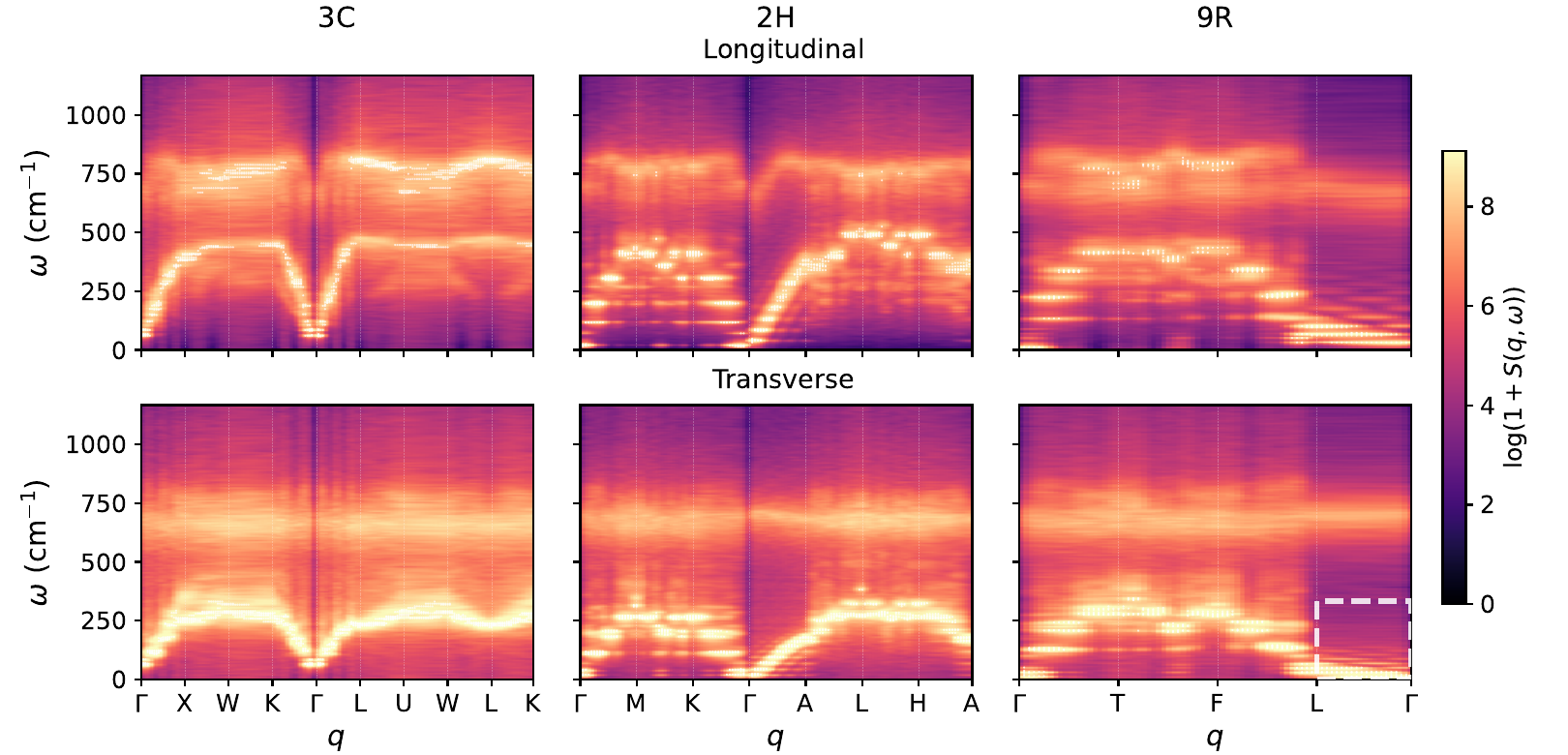}
    \caption{Longitudinal (top row) and transverse (bottom row) components of the dynamical structure factor $S(q,\omega)$ for the 3C, 2H, and 9R SiC polytypes at $T = 3000$~K. White markers denote peak positions extracted from $S(q,\omega)$ maxima. The dashed rectangle highlights the low-frequency shear-mode region in 9R.}
    \label{fig:phonons3}
\end{figure*}

\section{DFT functional selection and convergence testing}

\begin{figure*}[h!]
    \centering
    \includegraphics[width=0.95\textwidth]{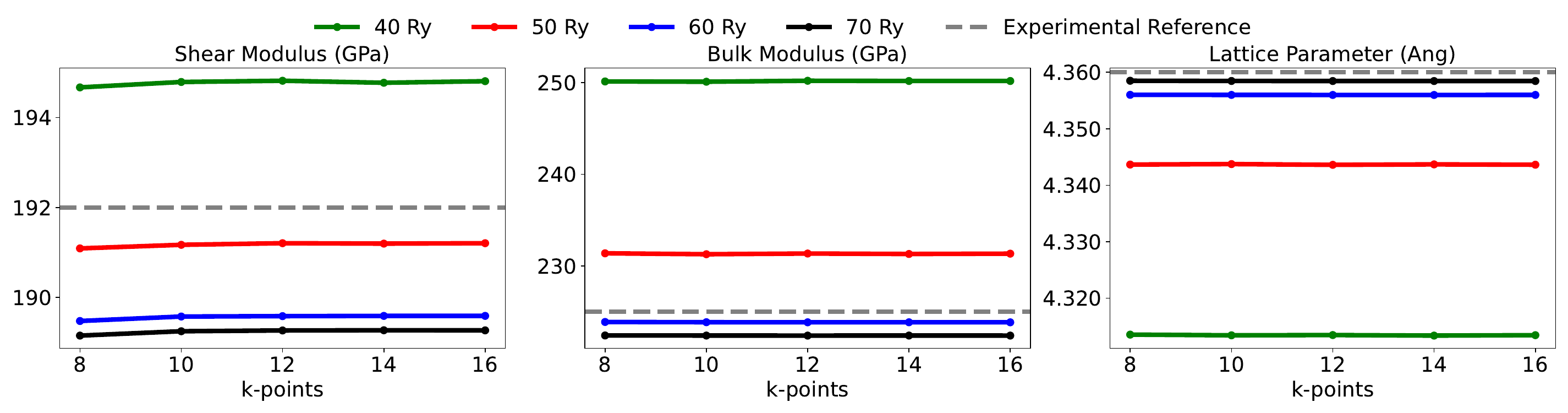}
    \caption{Tests for numerical convergence of DFT calculations on 3C-SiC using the PBEsol functional. Dashed lines show the reference values taken from experiments in the literature \cite{SiC_Elastic_JAmCeramSoc_1968,SiC_Elastic_PR_1968}.}
    \label{fig:ConvTest}
\end{figure*}

The basis set for our DFT calculations was determined following tests for numerical convergence. We chose as our measures of convergence the lattice parameter, the bulk modulus, and the shear modulus. Our target was to reach less than 1 GPa change in the moduli and less than 0.001~\AA~change in the lattice parameter. We investigated a range of plane-wave cutoff values and $k$-point grid densities (see Fig. \ref{fig:ConvTest}) and found that we achieve our convergence target at a basis set of 60 Ry plane-wave cutoff energy and a $12 \times 12 \times 12$ $k$-point grid.

We also compared the numerically converged predictions of 3C-SiC to published experimental data, shown in Table \ref{tab:convtest}. As PBEsol exhibited the closest match to measurements across our set of convergence measures, we chose to use the PBEsol functional for data generation.

\begin{table}
\begin{tabular}{r|ccc}
\hline
\hline
           & $a$~\AA & $B$ (GPa) & $G$ (GPa)\\
\hline
LDA        & 4.313   &  233.82   & 197.97 \\
PBE        & 4.381   &  213.38   & 188.43 \\
PBEsol     & 4.358   &  222.51   & 189.29 \\
Experiment & 4.36    &  225      & 192    \\
\hline
\hline
\end{tabular}
\caption{Comparison of numerically converged prediction of LDA, PBE, and PBEsol calculations
for the lattice parameter ($a$), bulk modulus ($B$), and shear modulus ($G$) to experimental
data \cite{SiC_Elastic_JAmCeramSoc_1968,SiC_Elastic_PR_1968} for the 3C phase of SiC.\label{tab:convtest}}
\end{table}

\newpage

\clearpage
\bibliography{references}

\end{document}